# Controllable suppression of Non-Hermitian skin effects

Chao Xu[1], Zhiqiang Guan[1,2,3,4*] and Hongxing Xu[1,2,4*]

[1] School of Physics and Technology, Center for Nanoscience and Nanotechnology, and Key Laboratory of Artificial Micro- and Nano-structures of Ministry of Education, Wuhan University, Wuhan 430072, China

[2] School of Microelectronics, Wuhan University, Wuhan 430072, China

[3] Hubei Yangtze Memory Laboratories, Wuhan 430205, China

[4] Wuhan Institute of Quantum Technology, Wuhan 430206, China

**Abstract**

The non-Hermitian skin effect (NHSE) is a phenomenon where the bulk states tend to the boundary within a non-Hermitian Hamiltonian system, with broad applications across various fields. A comprehensive understanding of anomalies in skin modes associated with NHSEs is essential for practical applications. Recently, some innovative works reported the suppression and enhancement of NHSEs through the application of magnetic fields, respectively. In our work, we engineered onsite potential energy distribution and found non-monotonic and monotonic suppression patterns on skin modes similar to magnetic fields. These suppression patterns represent characteristic transitions as the onsite potential distribution shifts from order to disorder. Relying only on onsite potential energy engineering, we have not only deepened our insight into the relationship and distinctions between order and disorder, but also developed a general strategy to demonstrate both robustness and controllable adjustability of the skin modes. By integrating with the scaling theory of disorder, we have extended the concept of controllable suppression of NHSEs to higher-dimensional systems.

**Introduction** --- In contrast to Hermitian systems, non-Hermitian systems are characterized by complex features such as non-real eigenvalues[1-5], non-orthogonal eigenvectors[6-8], and exceptional points (EPs) [9-16], which lead to a variety of intriguing phenomena. A significant and widespread property of non-Hermitian systems, as characterized by topological winding numbers, is the NHSE [17-25]. NHSE is a phenomenon in which a system exhibits bulk states localized at the boundary under open boundary conditions (OBCs) [26-31]. The necessary and sufficient conditions for non-zero spectral winding numbers with respect to a reference point can exist is that the energy eigenvalues form a closed curve with a certain interior on the energy complex plane under PBCs [32, 33]. In contrast to other topological invariants that are typically defined on the higher-dimensional or multi-band system, the winding numbers in this scenario are distinctively definable within a single energy band in one dimension. The ubiquitousness of the non-Hermitian effects and the low requirement for defining winding numbers, underscores the significant practical value of NHSEs. When there are higher dimensions, exhibiting more possible symmetries, or more complex systems, winding numbers will have richer connotations [34-37]. The NHSEs have been observed across diverse fields, including optics, acoustics, electric circuits, quantum walks, and synthetic dimensions [38-47].

Despite extensive theoretical and experimental studies on the NHSEs [38-50], research into the suppression or enhancement of skin modes and geometry-dependent skin effects is still relatively rare [32, 51-56]. Formulating a cohesive theoretical framework to explain these anomalies poses a significant challenge. From a theoretical standpoint, the genesis of skin modes can be attributed to two primary sources: engineering onsite potential energy and manipulation of hopping terms, while the vast majority of prior research focused has concentrated on the latter [33], our study shifts the focus to the former—onsite potential energy. Direct investigations into the role of potential energy are limited, yet the influence of various physical fields, such as electric [51] and magnetic fields [52, 56], can be translated into potential energy effects. These indirect studies have not provided a systematic understanding or fully elucidated the patterns of skin mode suppression. Recent innovative research has examined the influence of magnetic fields on NHSEs, with magnetic fields serving as a representation of potential energy. This research has uncovered a competitive dynamic between magnetic fields and non-Hermitian effects [52], highlighting the nonperturbative nature of magnetic fields, which tend to localize states within the bulk, contrasting with the skin-localizing tendency of non-Hermitian effects. An increase in the magnetic field intensity has been observed to suppress certain skin modes, driving them back into the bulk [52]. Additionally, there are also studies exploring the enhancement of second-order non-Hermitian skin effects via magnetic fields [56]. So there are two questions: how does the magnetic field act on the skin modes, and can we regulate them?

In this work, we constructed magnetic field-like onsite potential and random-like onsite potential, and found that onsite potential energy has non-monotonic and monotonic suppression patterns on skin modes, paving a novel path for regulating the skin modes. In the one-dimensional nearest-neighbor non-reciprocal hopping Hatano-Nelson model with periodic onsite potential energy, we first constructed a magnetic field-like potential energy distribution based on the Hofstadter model, then discovered that the suppression of skin mode initially increases and subsequently decreases with the increase of the magnetic field, simultaneously modulated by the lattice. From the perspective of electron tunneling through potential barriers, we speculate that there is a non-monotonic correlation between barrier disorder introduced by magnetic field and magnetic field intensity. To verify the hypothesis, we first need to extract the general relationship between (more general) barrier disorder and period as precisely as possible, because the magnetic field intensity is inversely proportional to the onsite potential period. Then, we constructed a random-like onsite potential energy distribution, which unveiled monotonic suppression patterns. Non-monotonic and monotonic suppression patterns are two typical forms for order and disorder onsite potential, respectively. Our exclusive focus on onsite potential energy not only facilitated a profound understanding of the relationship and distinctions between order and disorder, but also, we developed a general strategy to control skin mode suppression. Ultimately, by integrating our findings with the scaling theory of disorder, the concept of controllable skin mode suppression, based on the mutual transformation between order and disorder, is generalized to higher dimensions.

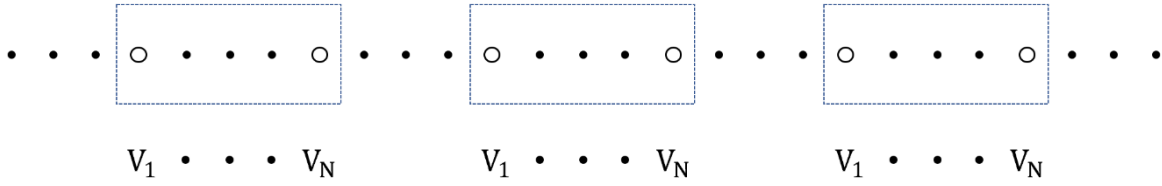

Fig.1 Schematic diagram of a one-dimensional single-atom chain considering onsite energies with period N. Dashed dots represent omitted atoms with potential energy $V_i$.

**Non-monotonic suppression by magnetic field ---** Our exploration of NHSEs commences with the fundamental one-dimensional one atom per site model. Non-Hermitian Hamiltonians often involve non-reciprocal hopping terms or complex onsite energy, introduced by gain or loss at sites. To maintain simplicity while ensuring representativeness, our initial analysis is confined to the most straightforward scenario. This involves considering solely the nearest-neighbor hoppings accompanied by a periodic arrangement of onsite potential energies, the Hamiltonian can be written as

$$\hat{H} = \sum_i t_{-1}\hat{a}_i^\dagger \hat{a}_{i+1} + t_{+1}\hat{a}_{i+1}^\dagger \hat{a}_i + \sum_i V_{i\,mod\,N}\hat{a}_i^\dagger \hat{a}_i \quad (1)$$

where $t_-, t_+$ are the hopping amplitude towards left and right sides, respectively. $V_{i\,mod\,N}$ is onsite potential energy at site *i* with period *N*, as shown in the schematic diagram in Fig. 1.

When examining a single electron within a non-reciprocal hopping square lattice subject to a vertical magnetic field, we encounter the renowned non-Hermitian Hofstadter model. The conventional Hermitian Hofstadter model is distinguished by its butterfly-shaped energy spectrum [57], a classical quantum fractal model in physics. It satisfies the Streda formula [58] and Wannier's Diophantine equation [59]. A key characteristic of the Hofstadter model is the possession of a magnetic translation group [60]. Employing the Peierls substitution [61] along the x-axis, the magnetic flux per unit cell can be denoted by $\phi = 2\pi \frac{p}{q}$, with *p, q* being coprime integers.

Here, $t_x$ represents the nearest-neighbor hopping amplitude in the x-axis direction, while $t_y^-$ and $t_y^+$ are the nearest-neighbor hopping amplitudes in the downward and upward directions of the y-axis, respectively. When the *x*-axis direction takes the PBCs, its Hamiltonian can be written as

$$\hat{H} = \sum_{k_x,n} \left( t_x e^{-ik_x+i\phi n}\hat{a}_n^\dagger \hat{a}_n + t_x e^{+ik_x-i\phi n}\hat{a}_n^\dagger \hat{a}_n + t_y^+ \hat{a}_{n+1}^\dagger \hat{a}_n + t_y^- \hat{a}_n^\dagger \hat{a}_{n+1} \right) \quad (2)$$

Non-reciprocal hopping terms do not break the magnetic translation group. Notably, the Hamiltonian at certain $k_x$ is equivalent to that of a one-dimensional single-atom chain considering onsite potential energies with period $q$, and their onsite potential distribution obeys a cosine function profile. The Hofstadter model serves as an important platform for investigating novel Hall effects [62]. Advances in theoretical physics have extended the concept of Chern numbers, traditionally associated with Hermitian Hamiltonians, to their non-Hermitian counterparts [29]. Consequently, the non-Hermitian Hofstadter model retains the capacity to define non-trivial Chern numbers, thereby supporting topological edge states. However, establishing an analytical quantification of the relationship between non-Hermitian strength and Chern numbers presents a formidable challenge, often due to the absence of an analytic solution within this model. (More information on the energy bands and state distribution of the Hofstadter model can be found in the Supplementary Materials, Sec. Ⅰ).

Setting $k_x = 0$ and considering a lattice with 500 atoms along the *y*-axis, the distribution of states is calculated in Fig. 2 as the ratio p/q increases. A judicious choice of simulation parameters ensures that topological edge states manifest exclusively on the right side. In Fig. 2(a), 2(b), and 2(c), we have highlighted the three states with the lowest energy in color. It is observable that in the absence of a magnetic field, the skin modes are concentrated to the boundary. Upon incrementing the magnetic field, certain skin modes penetrate the bulk, giving rise to what are termed anti-skin modes. Subsequently, by selecting a suitable magnetic field intensity, one, two and three wave packets in the spatial distribution are formed, as depicted in Fig. 2(c), 2(d), and 2(e). With a further increase in the magnetic field, the skin modes are observed to migrate back towards the boundary, as shown in Fig. 2(f). This behavior partially diverges from prior research which only reveals the exceptionally robust magnetic confinement effect that counteracts non-Hermiticity [52]. The popular physical image of the magnetic confinement effect is that of electrons becoming localized due to the Lorentz force within a magnetic field. However, we posit that this perspective may not be entirely accurate.

In our work, the magnetic field (intensity is proportional to $\frac{p}{q}$) can be regarded as a potential barrier that impedes electron hopping. In the absence of a magnetic field, electrons face no such barrier, allowing skin modes to typically accumulate at the boundary. However, as the magnetic field intensity augments, the elevated potential barrier obstructs electron hopping, resulting in skin modes moving into the bulk. Further increments in the magnetic field intensity lead to a reduction in the spatial period $\frac{q}{p}$ associated with the magnetic field. Provided the variance of the barrier distribution remains relatively stable, a shorter period facilitates electron tunneling, prompting the skin modes to gradually revert to the boundary. To validate our hypothesis, we introduced a triangular onsite potential distribution and observed analogous phenomena, reinforcing our understanding of the interplay between magnetic fields and skin modes. (see Supplementary Materials, Sec. Ⅲ)

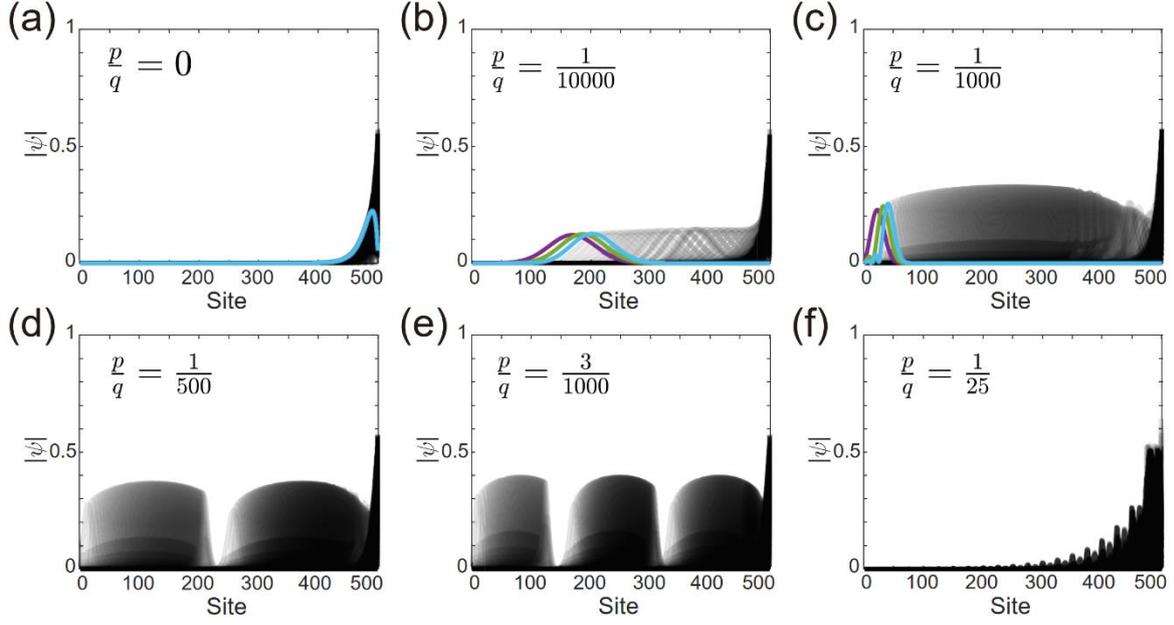

Fig.2 Skin modes suppression in Hofstadter model. Take $k_x = 0$, 500 atoms in the y-axis, and calculate state distribution under $t_x = 1, t_y^- = 0.9,$ $and\ t_y^+ = 1.1$. The three lowest energy states are marked in color in (a), (b), and (c).

**Monotonic suppression with random potential energy** --- In a finite chain, the modulation of skin modes is contingent upon the interplay between the disorder strength and the period of the onsite potential energy distribution. However, conventional potential distributions often struggle to independently manipulate these two factors due to their intrinsic coupling. To surmount this challenge and enable a more precise analysis, we have devised a cleaner system. Our approach involves employing a one-dimensional chain with a single atom per site, to which we introduce a periodic random onsite energy. This system allows for a greater degree of control in examining the relationship between disorder strength and period. Specifically, if the period $N$ is set equal to the total number $L$ of atoms in the chain, the model aligns with the non-Hermitian Anderson model [63-66]. Conversely, when the period $N$ is 1, the model simplifies to a non-Hermitian Hatano-Nelson model, which features a uniform constant onsite potential energy distribution.

We simulated the anti-skin mode proportion in a total number of atoms of $L = 400$, sampling 100 times under different periods $N$ and disorder amplitude $W$. It is found that the anti-skin modes are basically positively correlated as the period and disorder amplitude increases, as shown in Fig. 3 (More details about wavefunction distribution under disorder can be found in the Supplementary Materials, Sec. IV). Within a finite chain, if the disorder amplitude $W$ is fixed and the period $N$ is incrementally increased from 1 to the total number of atoms $L$, the system effectively transitions from a Hatano-Nelson model with constant potential energy to an Anderson model characterized by stronger localization effects. As the disorder strength escalates, it culminates in a rise in the proportion of anti-skin mode. Furthermore, when the period $N$ is fixed, and the disorder amplitude $W$ is incremented, the system mirrors the Anderson model, with the disorder strength increasing progressively. It is important to note that the inherent randomness in the onsite potential energy can introduce fluctuations in the proportion of anti-skin modes observed.

In contrast to the cosine or triangular onsite potential energy distributions previously considered, a random onsite potential energy distribution more effectively decouples the disorder strength from the period $N$. The disorder amplitude $W$ and the period $N$ are key factors that contribute to the overall disorder strength. Within the framework of the Anderson model, an increased disorder strength leads to a more localized state distribution. The principles governing Anderson localization, which addresses the effects of disorder on localization, apply similarly to non-random distributions, such as cosine distribution, triangular distribution, despite the differences in the nature of the disorder. Our understanding is reinforced by two main considerations. First, the Fourier transform allows us to conceive any periodic function as a linear combination of numerous sine or cosine functions. Second, extensive research has been dedicated to linking NHSEs with Anderson localization, extending the concept of a phase transition from extended to localized states in the Anderson model to the transition between skin modes and anti-skin modes in the context of NHSEs [67-69].

When there is a tighter coupling between the disorder strength and the period $N$, or when the potential barrier distribution approximates a simple functional form, the suppression of skin modes exhibits a characteristic non-monotonic pattern, or if not, monotonic pattern. Through the precise manipulation of the onsite potential energy distribution, a novel strategy for controlling the suppression pattern of skin modes emerges.

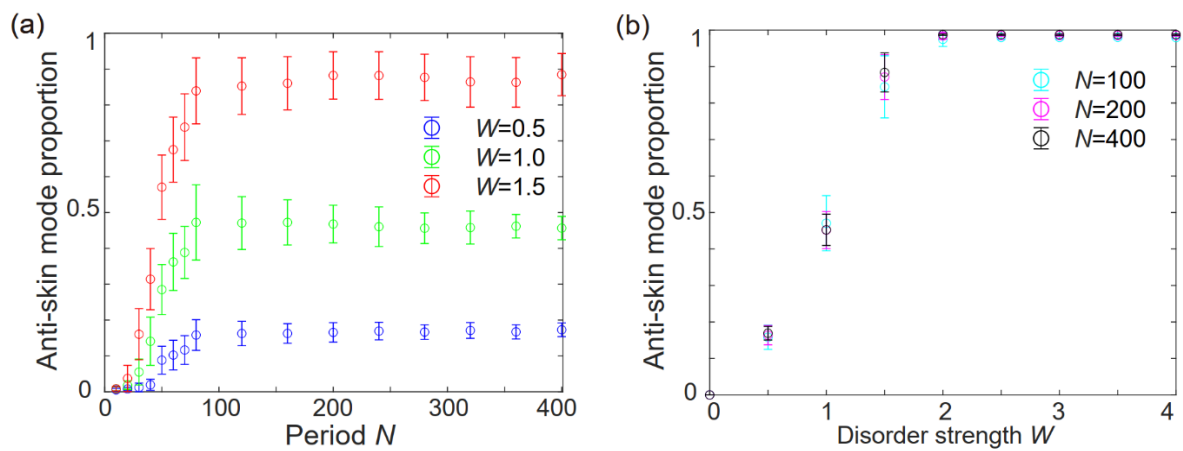

Fig.3 Anti-skin mode proportion in total modes under different period $N$ and disorder amplitude $W$. Parameters: $t_- = 0.9, t_+ = 1.1, V_{i\, mod\, N} \in [W, W], L = 400,$ sampling times is 100.

**Controllable suppression of skin modes in high dimensions** --- The NHSEs are intricately linked to the concept of Anderson localization, but due to the fact the onsite energy distribution is not necessarily random, more means of regulating the skin mode have emerged. When considering the two-dimensional skin mode, introducing a similar cosine onsite potential energy distribution as mentioned above, as shown in Fig.4, we can still observe that the skin mode exhibits a phenomenon of initially suppressing the skin modes, followed by a subsequent return to the boundary as the ratio $p/q$ escalates. The presence of two and three wave nodes in Fig .4(d) and 4(e), respectively, that are not located at the midpoint or one-third points of the chain can be attributed to boundary size effects. Extending our analysis to higher dimensions necessitates an engagement with the scaling theory pertinent to Anderson localization[64]. In the limit of infinitely large systems, all states in one-dimensional and two-dimensional systems are localized. However, a critical threshold exists for three-dimensional systems and beyond. When the level of disorder is subcritical, three-dimensional NHSEs may remain unsuppressed. Conversely, once the disorder surpasses this critical threshold, the NHSEs become entirely localized [64]. In scenarios involving a cosine distribution, the continuous augmentation of the cosine amplitude leads to an incremental rise in the overall disorder strength. Upon reaching the critical points, it becomes possible to suppress all skin modes. Thus, the NHSEs within three-dimensional systems remain subject to the constraints imposed by the scaling theory. The distinction lies in the capacity of the cosine function distribution to engender a more intricate interplay between disorder strength and period. This nuanced relationship facilitates a refined control over the suppression of skin modes, enabling the achievement of both robustness and tunability within these modes.

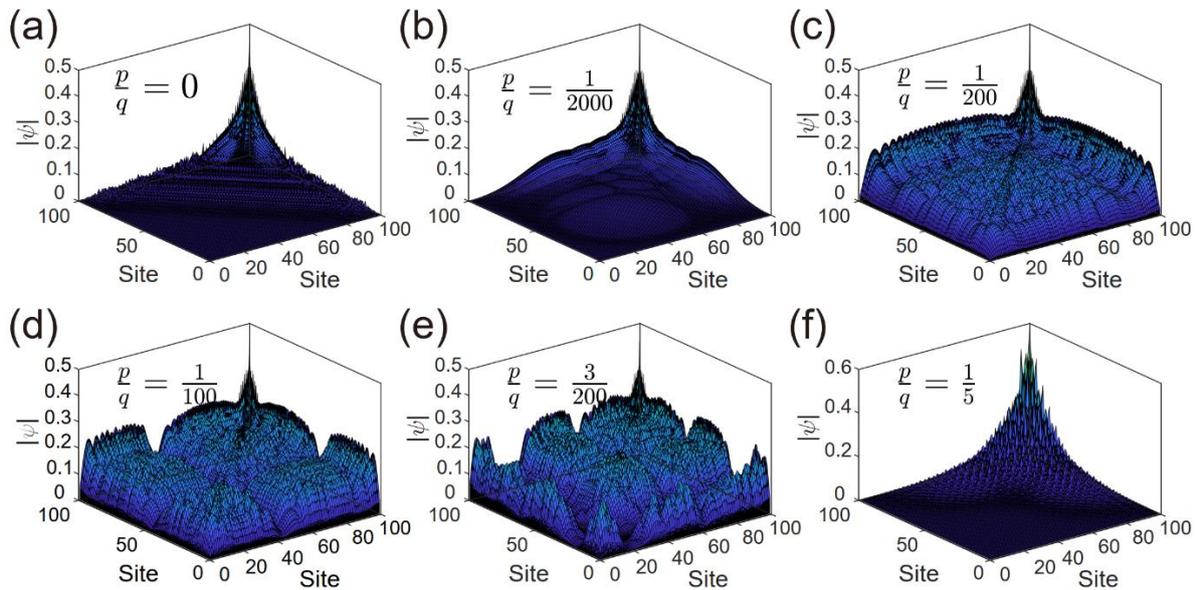

Fig.4 Skin modes suppression in two dimensions. Take 100 atoms with cosine onsite energy distribution applied along both x-axis and y-axis, and calculate state distribution under $t_x^- = 0.9, t_x^+ = 1.1, t_y^- = 0.9, t_y^+ = 1.1$.

**The conclusion and discussion ---** In this study, we initiated our investigation with the distribution of magnetic field-like onsite potential energy, uncovering non-monotonic suppression patterns exerted by magnetic fields on skin modes. To elucidate the origins of these patterns, we constructed a random-like potential energy distribution. Relying only on onsite potential energy engineering, we have not only deepened our insight into the relationship and distinctions between order and disorder, but also developed a general strategy to demonstrate both robustness and controllable adjustability of the skin modes.

The relationship between onsite potential energy and magnetic fields is characterized by a universal-special dichotomy, with the potential energy manifesting in various forms across different physical domains. Moreover, the concept of disorder extends beyond the realm of potential energy to encompass other areas, including the disorder in hopping terms. The suppression patterns induced by order and disorder can be extrapolated to more intricate scenarios, broadening our perspective on the underlying phenomena. Furthermore, the conventional framework of Generalized Brillouin Zones (GBZs) [70] fails to encapsulate the nuances of skin mode suppression. It is even plausible to postulate that all forms of localization may elude representation within GBZs [71]. This realization underscores the necessity for a more profound inquiry into the effects of localization on skin modes. The quest lies in identifying more efficacious methodologies to delineate and characterize these modes, propelling our research towards a more comprehensive grasp of the subject.

The work is supported by the National Key R&D Program of China (Grant No. 2021YFA1401100), the National Natural Science Foundation of China (No. 12174293), the Fundamental Research Funds for the Central Universities (No. 2042022kf0047), and the Special Fund of Hubei Yangtze Memory Laboratory.

∗Corresponding author.

zhiqiang.guan@whu.edu.cn

hxxu@whu.edu.cn

[1] Y. Choi, S. Kang, S. Lim, W. Kim, J.-R. Kim, J.-H. Lee and K. An, Phys. Rev. Lett. **104**, 153601 (2010)
[2] S. Diehl, E. Rico, M. A. Baranov and P. Zoller, Nature Phys. **7**, 971 (2011)
[3] E. Persson, I. Rotter, H. J. Stöckmann and M. Barth, Phys. Rev. Lett. **85**, 2478 (2000)
[4] P. Doyeux, S. A. H. Gangaraj, G. W. Hanson and M. Antezza, Phys. Rev. Lett. **119**, 173901 (2017)
[5] S. Buddhiraju, A. Song, G. T. Papadakis and S. Fan, Phys. Rev. Lett. **124**, 257403 (2020)
[6] W. Chen, Ş. Kaya Özdemir, G. Zhao, J. Wiersig and L. Yang, Nature **548**, 192 (2017)
[7] B. Peng, Ş. K. Özdemir, M. Liertzer, W. Chen, J. Kramer, H. Yılmaz, J. Wiersig, S. Rotter and L. Yang, Proc. Natl. Acad. Sci. **113**, 6845 (2016)
[8] J. Wiersig, Phys. Rev. Lett. **112**, 203901 (2014)
[9] A. A. Zyablovsky, A. P. Vinogradov, A. A. Pukhov, A. V. Dorofeenko and A. A. Lisyansky, Physics-Uspekhi **57**, 1063 (2014)
[10] Ş. K. Özdemir, S. Rotter, F. Nori and L. Yang, Nat. Mater. **18**, 783 (2019)
[11] V. V. Konotop, J. Yang and D. A. Zezyulin, Rev. Mod. Phys. **88**, 035002 (2016)
[12] L. Feng, R. El-Ganainy and L. Ge, Nat. Photonics **11**, 752 (2017)
[13] R. El-Ganainy, K. G. Makris, M. Khajavikhan, Z. H. Musslimani, S. Rotter and D. N. Christodoulides, Nature Phys. **14**, 11 (2018)
[14] J. Wiersig, Phys. Rev. A **93**, 033809 (2016)
[15] J. Zhu, S. K. Ozdemir, Y.-F. Xiao, L. Li, L. He, D.-R. Chen and L. Yang, Nat. Photonics **4**, 122 (2010)
[16] S. Dong, G. Hu, Q. Wang, Y. Jia, Q. Zhang, G. Cao, J. Wang, S. Chen, D. Fan, W. Jiang, Y. Li, A. Alù and C.-W. Qiu, ACS Photonics **7**, 3321 (2020)

[17] T. Liu, Y.-R. Zhang, Q. Ai, Z. Gong, K. Kawabata, M. Ueda and F. Nori, Phys. Rev. Lett. **122,** 076801 (2019)
[18] C. H. Lee, L. Li and J. Gong, Phys. Rev. Lett. **123,** 016805 (2019)
[19] D. S. Borgnia, A. J. Kruchkov and R.-J. Slager, Phys. Rev. Lett. **124,** 056802 (2020)
[20] K. Kawabata, T. Bessho and M. Sato, Phys. Rev. Lett. **123,** 066405 (2019)
[21] F. K. Kunst, E. Edvardsson, J. C. Budich and E. J. Bergholtz, Phys. Rev. Lett. **121,** 026808 (2018)
[22] F. Song, S. Yao and Z. Wang, Phys. Rev. Lett. **123,** 170401 (2019)
[23] S. Yao, F. Song and Z. Wang, Phys. Rev. Lett. **121,** 136802 (2018)
[24] S. Yao and Z. Wang, Phys. Rev. Lett. **121,** 086803 (2018)
[25] N. Okuma, K. Kawabata, K. Shiozaki and M. Sato, Phys. Rev. Lett. **124,** 086801 (2020)
[26] C.-K. Chiu, J. C. Y. Teo, A. P. Schnyder and S. Ryu, Rev. Mod. Phys. **88,** 035005 (2016)
[27] M. Z. Hasan and C. L. Kane, Rev. Mod. Phys. **82,** 3045 (2010)
[28] X.-L. Qi and S.-C. Zhang, Rev. Mod. Phys. **83,** 1057 (2011)
[29] H. Shen, B. Zhen and L. Fu, Phys. Rev. Lett. **120,** 146402 (2018)
[30] N. P. Armitage, E. J. Mele and A. Vishwanath, Rev. Mod. Phys. **90,** 015001 (2018)
[31] E. J. Bergholtz, J. C. Budich and F. K. Kunst, Rev. Mod. Phys. **93,** 015005 (2021)
[32] K. Zhang, Z. Yang and C. Fang, Nat. Commun. **13,** 2496 (2022)
[33] K. Zhang, Z. Yang and C. Fang, Phys. Rev. Lett. **125,** 126402 (2020)
[34] Z. Gong, Y. Ashida, K. Kawabata, K. Takasan, S. Higashikawa and M. Ueda, Phys. Rev. X **8,** 031079 (2018)
[35] K. Kawabata, K. Shiozaki, M. Ueda and M. Sato, Phys. Rev. X **9,** 041015 (2019)
[36] D. Leykam, K. Y. Bliokh, C. Huang, Y. D. Chong and F. Nori, Phys. Rev. Lett. **118,** 040401 (2017)
[37] Y. Xiong, Journal of Physics Communications **2,** 035043 (2018)
[38] M. Brandenbourger, X. Locsin, E. Lerner and C. Coulais, Nat. Commun. **10,** 4608 (2019)
[39] T. Hofmann, T. Helbig, F. Schindler, N. Salgo, M. Brzezińska, M. Greiter, T. Kiessling, D. Wolf, A. Vollhardt, A. Kabaši, C. H. Lee, A. Bilušić, R. Thomale and T. Neupert, Phys. Rev. Research **2,** 023265 (2020)
[40] A. Ghatak, M. Brandenbourger, J. Van Wezel and C. Coulais, Proc. Natl. Acad. Sci. **117,** 29561 (2020)
[41] K. Wang, A. Dutt, K. Y. Yang, C. C. Wojcik, J. Vučković and S. Fan, Science **371,** 1240 (2021)
[42] H. Xin, W. Song, S. Wu, Z. Lin, S. Zhu and T. Li, Phys. Rev. B **107,** 165401 (2023)
[43] Z. Gu, H. Gao, H. Xue, J. Li, Z. Su and J. Zhu, Nat. Commun. **13,** 7668 (2022)
[44] L. Zhang, Y. Yang, Y. Ge, Y.-J. Guan, Q. Chen, Q. Yan, F. Chen, R. Xi, Y. Li, D. Jia, S.-Q. Yuan, H.-X. Sun, H. Chen and B. Zhang, Nat. Commun. **12,** 6297 (2021)
[45] D. Zou, T. Chen, W. He, J. Bao, C. H. Lee, H. Sun and X. Zhang, Nat. Commun. **12,** 7201 (2021)
[46] L. Xiao, T. Deng, K. Wang, G. Zhu, Z. Wang, W. Yi and P. Xue, Nat. Phys. **16,** 761 (2020)
[47] M.-J. Liao, M.-S. Wei, Z. Lin, J. Xu and Y. Yang, Results in Physics **57,** 107372 (2024)
[48] X.-R. Ma, K. Cao, X.-R. Wang, Z. Wei, Q. Du and S.-P. Kou, Phys. Rev.Research **6,** 013213 (2024)
[49] X. Zhang, T. Zhang, M.-H. Lu and Y.-F. Chen, Advances in Physics: X **7,** 2109431 (2022)
[50] S.-Q. Wu, Y. Xu and J.-H. Jiang, Frontiers in Physics **11,** 1123596 (2023)
[51] Y. Peng, J. Jie, D. Yu and Y. Wang, Phys. Rev. B **106,** L161402 (2022)
[52] M. Lu, X.-X. Zhang and M. Franz, Phys. Rev. Lett. **127,** 256402 (2021)
[53] C.-X. Guo, C.-H. Liu, X.-M. Zhao, Y. Liu and S. Chen, Phys. Rev. Lett. **127,** 116801 (2021)
[54] Z. Fang, M. Hu, L. Zhou and K. Ding, Nanophotonics **11,** 3447 (2022)
[55] Y.-C. Wang, J.-S. You and H. H. Jen, Nat. Commun. **13,** 4598 (2022)
[56] C.-A. Li, B. Trauzettel, T. Neupert and S.-B. Zhang, Phys. Rev. Lett. **131,** 116601 (2023)
[57] D. R. Hofstadter, Phys. Rev. B **14,** 2239 (1976)
[58] P. Streda, J. Phys. C: Solid State Phys. **15,** L717 (1982)

[59] G. H. Wannier, Phys. Status Solidi (B) **88,** 757 (2006)
[60] J. Zak, Phys. Rev. **134,** A1602 (1964)
[61] R. Peierls, Zeitschrift fr Physik **80,** 763 (1933)
[62] D. J. Thouless, M. Kohmoto, M. P. Nightingale and M. Den Nijs, Phys. Rev. Lett. **49,** 405 (1982)
[63] P. W. Anderson, Phys. Rev. **109,** 1492 (1958)
[64] E. Abrahams, P. W. Anderson, D. C. Licciardello and T. V. Ramakrishnan, Phys. Rev. Lett. **42,** 673 (1979)
[65] D. J. Thouless, J. Phys. C: Solid State Phys. **3,** 1559 (1970)
[66] N. Mott, J. Phys. C: Solid State Phys. **20,** 3075 (1987)
[67] H. Jiang, L.-J. Lang, C. Yang, S.-L. Zhu and S. Chen, Phys. Rev. B **100,** 054301 (2019)
[68] J. Claes and T. L. Hughes, Phys. Rev. B **103,** L140201 (2021)
[69] R. Sarkar, S. S. Hegde and A. Narayan, Phys. Rev. B **106,** 014207 (2022)
[70] K. Yokomizo and S. Murakami, Phys. Rev. Lett. **123,** 066404 (2019)
[71] R. Abou-Chacra, D. J. Thouless and P. W. Anderson, J. Phys. C: Solid State Phys. **6,** 1734 (1973)

# Supplemental Material for " A general strategy to control suppression of Non-Hermitian skin effects "

Chao Xu[1], Zhiqiang Guan[1,2,3,4*] and Hongxing Xu[1,2,4*]

[1] School of Physics and Technology, Center for Nanoscience and Nanotechnology, and Key Laboratory of Artificial Micro- and Nano-structures of Ministry of Education, Wuhan University, Wuhan 430072, China

[2] School of Microelectronics, Wuhan University, Wuhan 430072, China

[3] Hubei Yangtze Memory Laboratories, Wuhan 430205, China

[4] Wuhan Institute of Quantum Technology, Wuhan 430206, China

## I Non-Hermitian Hofstadter model

The Hofstadter model originates from the challenge of determining the energy spectrum for a single electron within a square lattice subjected to a perpendicular magnetic field. A defining characteristic of this model is its incorporation of a magnetic translation group. The Peierls substitution is implemented in the direction of the x-axis, with the magnetic flux through each unit cell expressed as $\phi = 2\pi\frac{p}{q}$, where p, q are coprime integers. The x-axis nearest-neighbor hopping amplitude is denoted by $t_x$, while $t_y^-$ and $t_y^+$ represent the nearest-neighbor hopping amplitudes in the downward and upward directions of the y-axis, respectively. Consequently, the Hamiltonian for the system can be formulated as follows:

$$\hat{H} = \sum_{m,n} (t_x \hat{c}_{m+1,n}^\dagger \hat{c}_{m,n} e^{i\phi n} + t_x \hat{c}_{m,n}^\dagger \hat{c}_{m+1,n} e^{-i\phi n} + t_y^+ \hat{c}_{m,n+1}^\dagger \hat{c}_{m,n} + t_y^- \hat{c}_{m,n}^\dagger \hat{c}_{m,n+1}) \quad (S1)$$

Non-reciprocal hoppings do not break the magnetic translation group. When reciprocal nearest-neighbor hoppings are assumed along the y-axis, that is, when $t_y^- = t_y^+$, the system exhibits *q* zero-energy Dirac points. These points are protected by a hidden chiral symmetry, a condition that is met when *q* is an even integer. In our selection of parameters, both Hermitian and non-Hermitian systems are capable of giving rise to topological edge states under open boundary conditions (OBCs) applied to the y-axis. Notably, for the non-Hermitian system, the bulk states are observed to converge towards the edges, as depicted in Fig. 1(d).

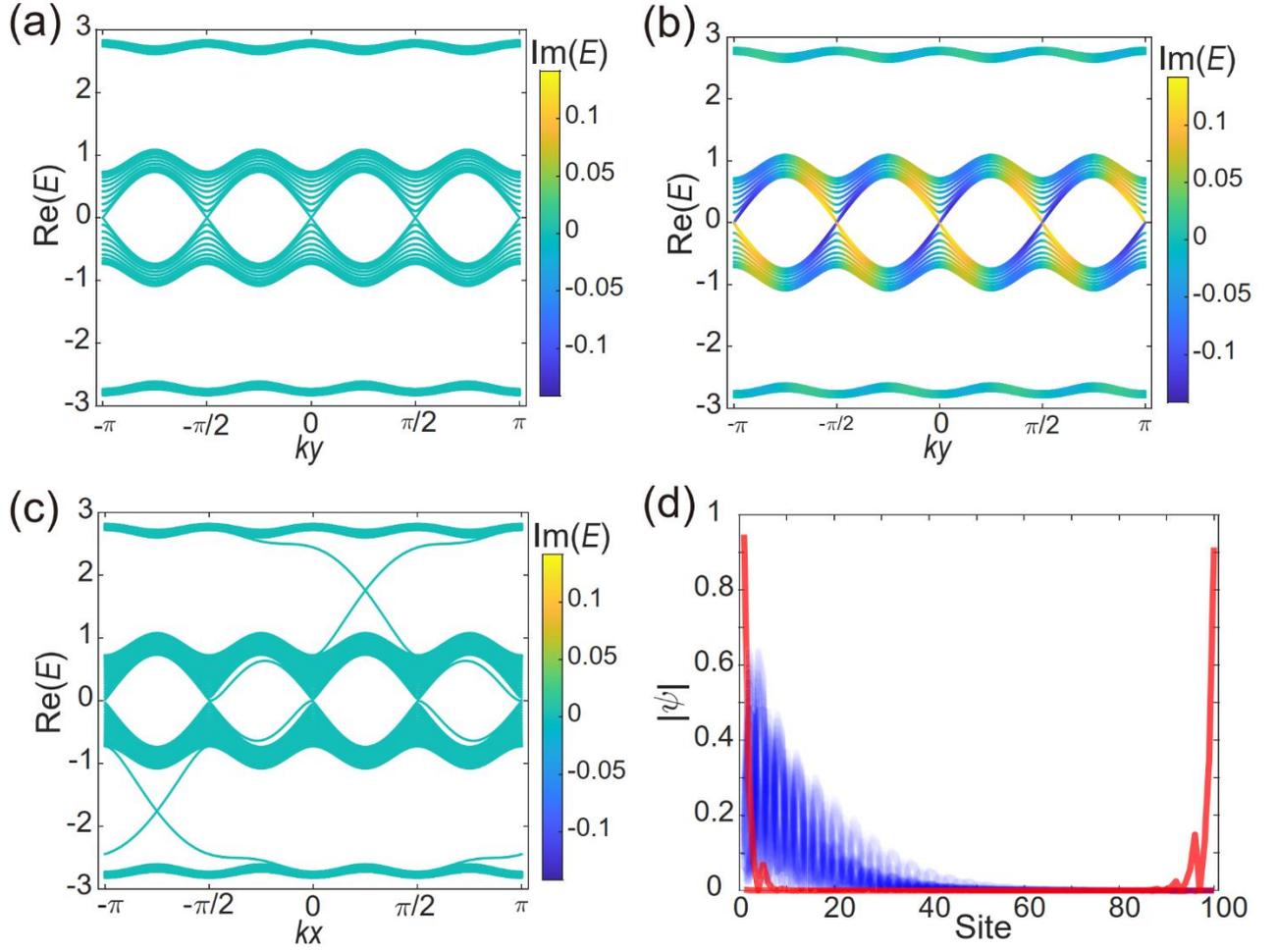

Fig. 1 Non-Hermitian Hofstadter model. Panel (a), (b) are the projected energy bands for $p=1$, $q=4$ taking periodic boundary conditions at both directions, where (a), (b) nearest neighbor hoppings are $t_x = t_y^- = t_y^+ = 1$ and $t_x = 1, t_y^- = 1.1, t_y^+ = 0.9$, respectively. (c) is the projected energy band along the x-axis direction, and the y-direction takes open boundary conditions (100 atoms in the y-axis). (d) is the distribution of bulk states (in black) and edge states (in red) in (c) for $k_x = \frac{\pi}{4}$.

## II The suppression of non-Hermitian skin effects in the Hofstadter model

Hermitian Hofstadter models are renowned for possessing non-trivial Chern numbers. These models retain their non-trivial Chern numbers even when the nearest-neighbor hoppings are made non-reciprocal. Fig. 1(d) illustrates the state distribution, which encompasses both the topological edge states and the skin modes. Subsequent analysis focuses on the state distribution across various values of $p$ and $q$ for the magnetic flux $\phi =$

$2\pi\frac{p}{q}$. Employing the Hamiltonian outlined in eq. 2 of the main text, with the stipulation that $t_x = t_y^- = t_y^+ = 1$, and considering a system with 500 atoms along the y-axis under open boundary conditions (OBCs), we have computed the projected energy bands for diverse combinations of p and q in the x-axis direction.

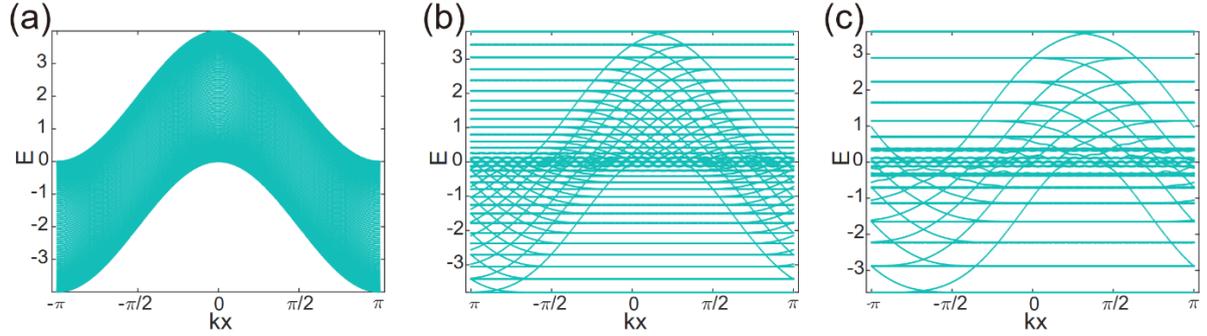

Fig.2 Projected energy bands of the two-dimensional Hermitian Hofstadter model under the OBCs (number of atoms L=500) along the *y*-axis direction and PBCs along the *x*-axis direction for different magnetic fluxes. *p* and *q* are $\frac{p}{q} = \frac{0}{31}$, $\frac{p}{q} = \frac{1}{31}$, $\frac{p}{q} = \frac{2}{31}$ in (a), (b), and (c), respectively.

A judicious selection of simulation parameters ensures that topological edge states manifest exclusively on a single side of the system. With setting $k_x = 0$, we proceed to calculate the state distribution under two distinct sets of parameters: first, with $t_x = t_y^- = t_y^+ = 1$ and second, with $t_x = 1, t_y^- = 0.9, t_y^+ = 1.1$, respectively.

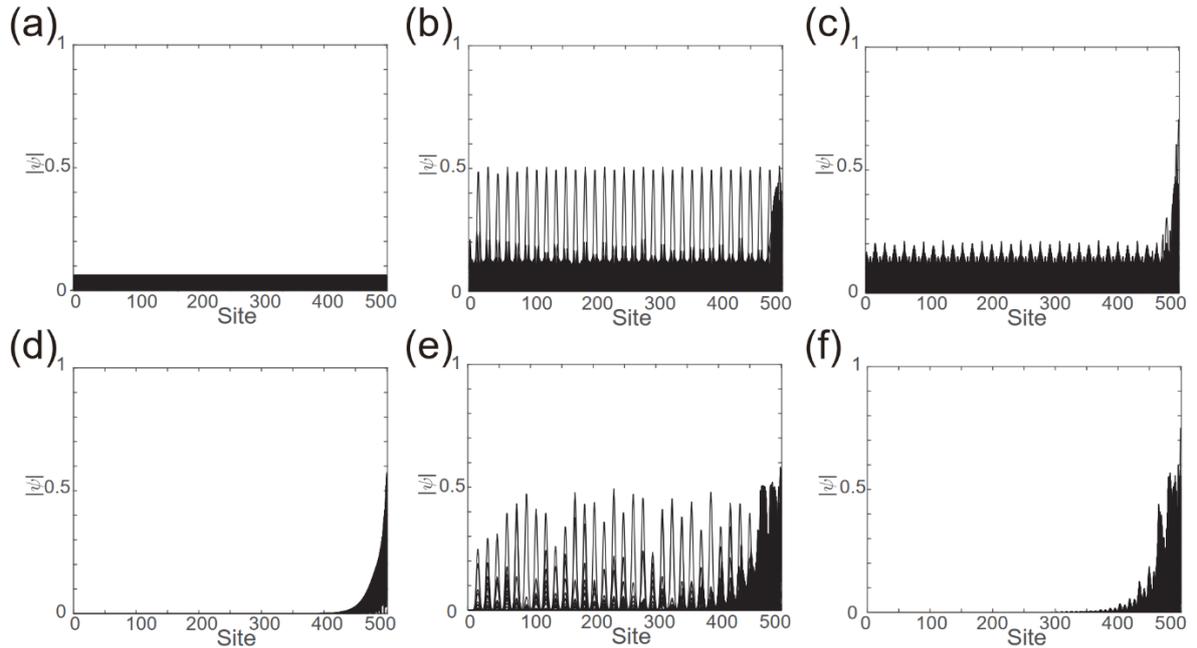

Fig.3 State distribution in Hofstadter model under OBCs with different magnetic field intensities. Take $k_x = 0$ and calculate state distribution under $t_x = t_y^- = t_y^+ = 1$ in (a), (b), (c) and $t_x = 1, t_y^- = 0.9, t_y^+ = 1.1$ in (d), (e), (f), respectively. The parameters of *p, q* in (a), (d) are $\frac{p}{q} = \frac{0}{31}$, ones in (b), (e) are $\frac{p}{q} = \frac{1}{31}$, ones in (c), (f) are $\frac{p}{q} = \frac{2}{31}$.

Fig. 3 shows that it is possible to generate topological edge states in the system when adding a magnetic field. If non-reciprocal hopping strength is considered, the concentration degree of the topological edge states will also be affected to a certain extent. More intriguingly, focusing on the bulk states reveals that in the absence of a magnetic field, the skin modes possess a minimal skin depth. Interestingly, the introduction of a weak magnetic field can prompt some skin modes to retreat back into the bulk. Conversely, when the magnetic field is intensified, these skin modes are drawn towards the boundary. The observed suppression pattern of the skin modes with varying magnetic field strengths is quite unexpected and merits further investigation.

## III Skin modes anomaly in trigonometric onsite energy distribution

As previously discussed, the projected energy band state distribution of the Non-Hermitian Hofstadter model, when subjected to a magnetic field, is analogous to that of a one-dimensional monoatomic chain with cosine-distributed onsite energy. To explore further, we introduce a trigonometric variation to the onsite energy distribution. Specifically, if the site number $i$ meets the condition $0 \leq \{i * \frac{p}{q}\} \leq \frac{1}{2}$, we define $V_i = \{i * \frac{p}{q}\}$; if $\frac{1}{2} < \{i * \frac{p}{q}\} < 1$, we set $V_i = 1 - \{i * \frac{p}{q}\}$. Here, $\{i * \frac{p}{q}\}$ represents the fractional part of $i * \frac{p}{q}$. Following the same approach, we select an open chain comprising 500 atoms and compute the state distribution as $p/q$ increases, as depicted in Fig. 4. The behavior of the skin modes, which initially suppress and then re-emerge, parallels that observed with cosine distribution. This observation underscores the intricate interplay between the period and the degree of disorder, a relationship that is not exclusive to magnetic fields.

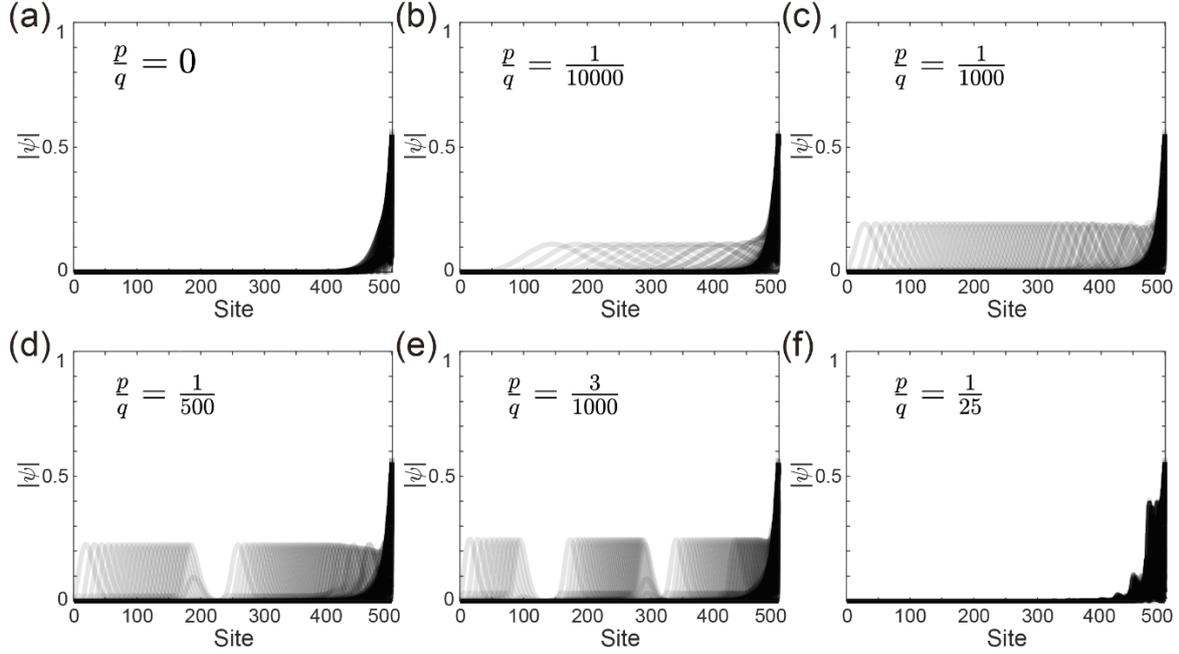

Fig.4 Skin modes anomaly in trigonometric onsite energy distribution. Take 500 atoms in a 1D monoatomic chain and calculate state distribution under $t_- = 0.9, t_+ = 1.1$.

## IV Wavefunction distribution under disorder

In a one-dimensional atomic chain, if only the nearest neighbor non-reciprocal hoppings are considered, the bulk state will be localized to the boundary. When such a chain is perturbed by external influences or additional onsite energy, the bulk states, protected by topological invariants, are drawn towards the boundaries as well. However, when the imposed disorder surpasses a certain threshold, the bulk states can revert to the bulk region due to Anderson localization. While the main text has discussed two variables associated with the disorder, a more detailed depiction of the wavefunction distribution has not been provided. To address this, we have employed a consistent methodology to assess the influence of varying the period $N$ on the wavefunction, ensuring that the variance of the random onsite energy distribution is conserved as possible. As illustrated in Fig. 5, for a chain comprising 400 atoms in total ($L$=400), there is an observable trend: an increase in the period $N$ correlates with a heightened degree of disorder. The anti-skin mode, influenced by pronounced boundary size effects, does not manifest a strictly periodic spatial distribution. Nonetheless, when the calculation is extended to a chain with 2000 atoms in total, the spatial arrangement of the skin mode becomes more apparent, exhibiting a more orderly pattern. Concurrently, the prevalence of anti-skin modes escalates with an increase in the period $N$.

Let's envision a thought experiment that explores the extreme cases of disorder in a one-dimensional system by considering two limiting scenarios: $N$=1 and $N$=$L$. The first case, where $N$=1, corresponds to the Hatano-Nelson model characterized by a constant potential energy across the lattice sites. Conversely, when N=L, the system aligns with the traditional Anderson model, which is defined by a disordered potential. As N increases from 1 to L, the system effectively transitions from one model to another, embodying a spectrum of disorder strengths. Notably, the strength of disorder in this transitional process is positively correlated with period N, offering a continuity between the ordered Hatano-Nelson and the disordered Anderson regimes.

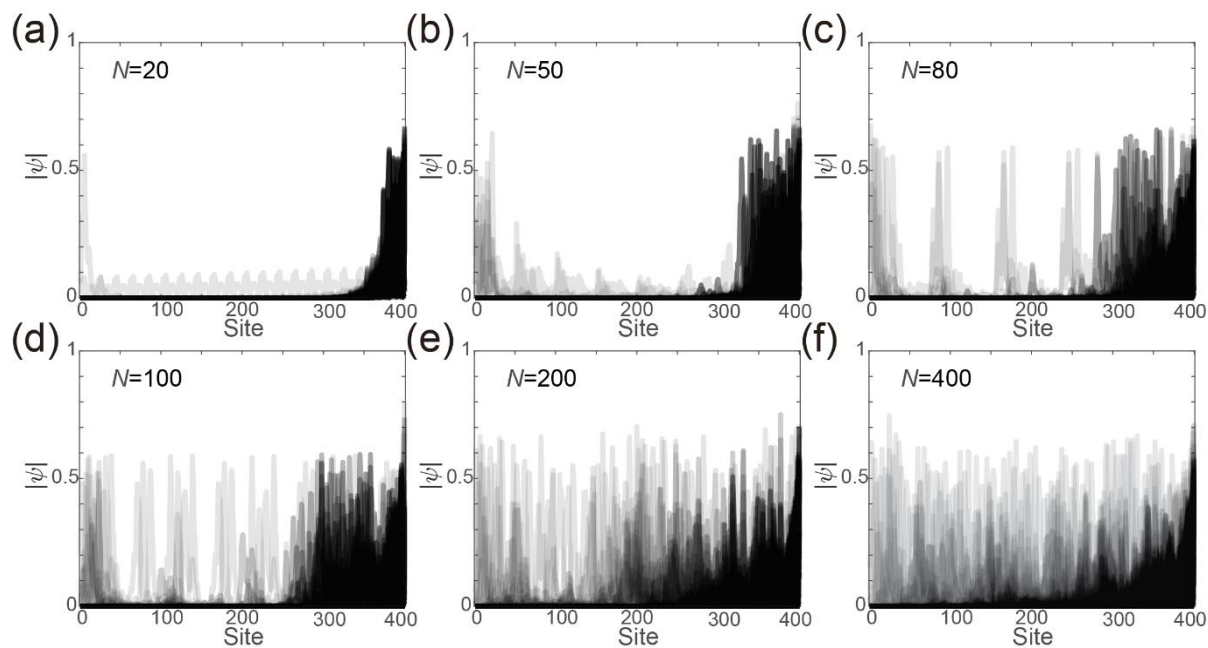

Fig.5 Wavefunction distribution under different period N with whole site number $L = 400$. Parameters: $t_- = 0.9, t_+ = 1.1, V_{i\,mod\,N} \in [W, W], W = 1$.

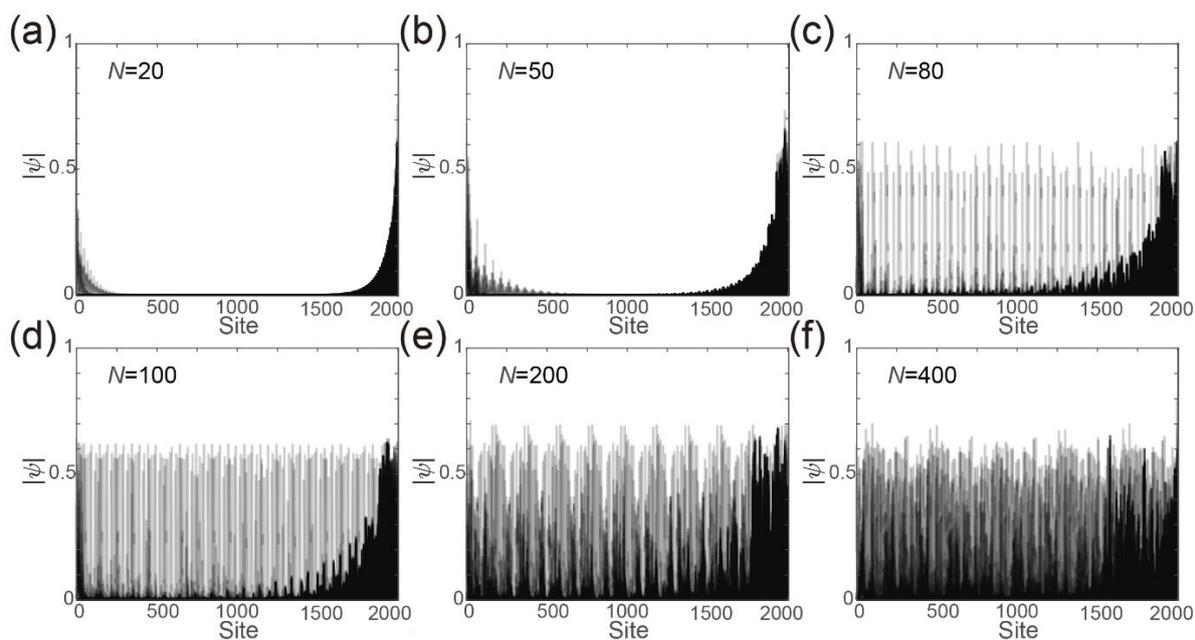

Fig.6 Wavefunction distribution under different period N whole site number $L = 2000$. Parameters: $t_- = 0.9, t_+ = 1.1, V_{i\,mod\,N} \in [W, W], W = 1$.

We then continued to investigate the impact of varying the amplitudes of the random onsite energy distribution. For this purpose, we selected a chain consisting of 1000 atoms and imposed periodic boundary conditions to plot the energy bands on the complex plane. Observing the system as the disorder increases, we noted that the enclosed space within the energy band progressively diminishes until it vanishes at a critical point, a phenomenon known as the Anderson transition. This critical transition, indicative of a shift from extended to localized states, is not predictable through the Generalized Brillouin Zones. Subsequently, under open boundary conditions, we delineated the state distribution for a chain of 100 atoms across three distinct levels of disorder intensity. Our findings reveal the gradual retraction of the skin mode back into the bulk, illustrating the nuanced interplay between disorder and skin modes.

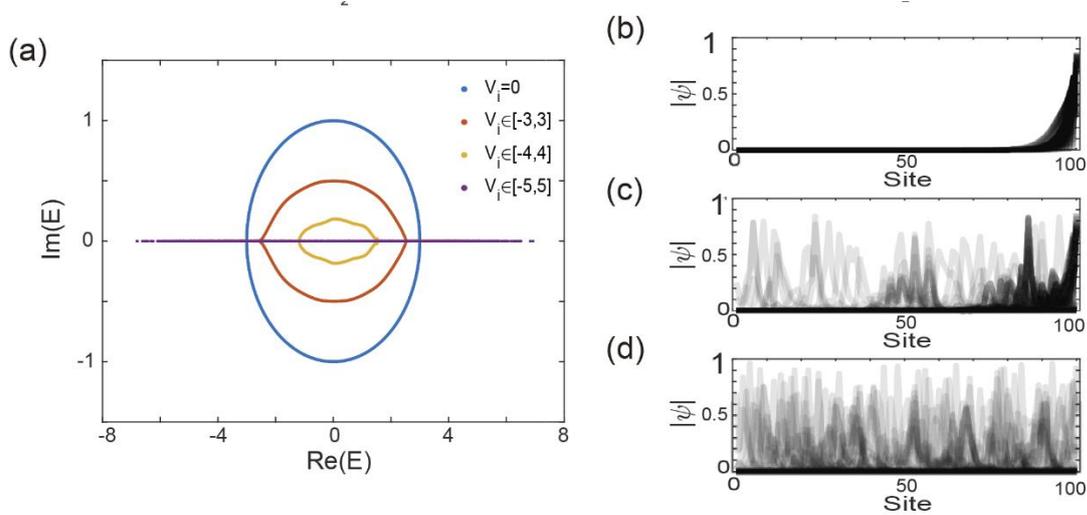

Fig.7 Wavefunction distribution under different disorder amplitude. Panel (a) is the complex spectra with parameters $t_- = 1$, $t_+ = 2$, subject to random real onsite energies under the periodic boundary conditions (atom numbers L=1000), $V_i$ is the random onsite energies on site *i*. Panel (b), (c), d) are state distribution when $V_i = 0, V_i \in [-3,3], V_i \in [-5,5]$ in panel (c).